\documentclass{jpsj3}

\title{Anisotropy of   $\pi$-plasmon Dispersion Relation of AA-stacked Graphite}

\author{Chih-Wei Chiu, Feng-Lin Shyu$^1$\thanks{E-mail address: fl.shyu@msa.hinet.net}, Ming-Fa Lin\thanks{E-mail address: mflin@mail.ncku.edu.tw}, Godfrey Gumbs$^2$\thanks{E-mail address: ggumbs@hunter.cuny.edu}, Oleksiy Roslyak$^2$ 
}
\inst{Department of Physics, National Cheng Kung University, 701 Tainan, Taiwan \\
$^{1}$Department of Physics, R.O.C. Military Academy, 830 Kaohsiung, Taiwan \\
$^{2}$Department of Physics and Astronomy, Hunter College at the City University of New York, 695 Park Avenue New York, NY 10065 USA}
\abst{The dispersion relation of the high energy optical $\pi$-plasmons  of simple hexagonal intrinsic graphite was calculated within the self-consistent-field approximation. The plasmon frequency $\omega_p$ is determined as functions of the transferred momentum $q_\parallel$ along the hexagonal plane in the
Brillouin zone and its perpendicular component $q_z$. These plasmons  are isotropic within the plane in the long wavelength limit. As the in-plane transferred momentum is increased, the plasmon frequency strongly depends on its magnitude and direction ($\phi$). With increasing angle, the dispersion relation within the hexagonal plane is gradually changed from quadratic to nearly linear form. There are many significant differences for the $\pi$-plasmon dispersion relations between 2D graphene and 3D AA-stacked graphite. They include   $q_\parallel$- and $\phi$-dependence and  $\pi$-plasmon bandwidth. This result reveals that interlayer interaction could enhance anisotropy of in-plane $\pi$-plasmons. For chosen $\textbf{q}_\parallel$,  we also obtain the plasmon frequency as a function of $q_z$ and show that there is an upper bound on $q_z$ for plasmons to exist in graphite. Additionally, the group velocity for plasmon propagation along the perpendicular direction  may be positive or negative depending on the choice of $\textbf{q}_\parallel$. Consequently, the forward and backward propagation of $\pi$-plasmons in  AA-stacked graphite in which the energy flow is respectively parallel or antiparallel to the transferred momentum, can be realized. The backward flowing resonance is an intrinsic property of AA-stacked graphite, arising from the energy band structure and the interlayer coupling.}

\kword{$\pi$-plasmon, AA-stacked graphite, anisotropy}

\begin{document}
\maketitle

\section{Introduction}

Monolayer graphene, a two-dimensional material purely made of carbon atoms arranged in a hexagonal lattice, was created on top of a SiO$_2$/Si substrate.$^{1)}$
Its low-energy electronic  structure shows linear dispersion corresponding to the massless Dirac fermion. The density of states vanishes at Fermi energy that leads to a zero-gap
semiconductor. Such pure 2D system are identified to display rich
physical properties, e.g., the peculiar Landau level quantization,$^{2)}$ and the novel half-integer quantum Hall effect.$^{3)}$

When a series of parallel graphene planes are periodically stacked, the 3D graphite crystal is formed. Graphite is one of the most important layered systems and has attracted many theoretical$^{4-26)}$ and experimental$^{27-34)}$ studies.
In terms of stacking types, there are four kinds of layered graphites. They include the simple hexagonal graphite (AA-stacked graphite),$^{12,15,16)}$ the Bernal graphite (AB-stacked graphite),$^{12,16,20)}$ the rhombohedral graphite
(ABC-stacked graphite),$^{13,17,21)}$ and the turbostratic graphite (without the periodic stacking sequence).$^{13)}$
The AB-stacked graphite is the most common and stable layered graphite. The ABC-stacked graphite naturally exists just in combination with the AB-stacked graphite.  It also has been synthesized in lab.$^{20,21)}$ As for the AA-stacking sequence, it only existed in the Li-intercalated graphite before, but was  synthesized in experiments recently.$^{29-31)}$ By the way, there were experiments to stress the influences of the stacking effect on physical properties. For examples: (1) angle-resolved photoemission (ARPES) was used to study electron-electron correlations,$^{32)}$ (2) high-resolution ARPES was performed to study quasiparticle
dynamics,$^{33)}$ and (3) the measurements of inelastic X-ray scattering on the spectra near Bragg reflections.$^{34)}$

Plasmon, its quantity measured in electron-energy-loss spectroscopy (EELS) is the loss function, which is given by
$Im[-1/\epsilon(q,\omega)]$. This measurement provides information on the collective excitations of the electronic system. The loss function essentially describes plasmon excitations that arise from intraband and interband transitions. In contrast to optical spectroscopy, which probes only dipole-allowed ($q=0$) vertical transition, one gets access here to the momentum-dependence of the dielectric function, i.e., non-vertical transitions. Consequently, high-resolution EELS could be used to study important quantities such as the
character of excitonic excitations, the spatial extension of excitons, and the degree of localization of electronic intraband and interband excitations. Investigations on plasmons, except for understanding the elementary excitations, there have been recent studies involving nanoplasmonics of graphene.$^{35)}$ The gapless manetoplasmon was obtained by controlling the external electric fields, that could be used to develop plasmon-enhanced imaging and sensing techniques.$^{36)}$ Another possible utilization of plasmons is applied to integrated optical circuits.

The $\pi$-plasmons in  monolayer and muitilayer graphene and graphite had been measured by  EELS$^{37-40)}$ or optical spectra.$^{41-44)}$ The $\pi$-plasmon is identified from the most pronounced peak in the loss function. The experimental measurements show that the $\pi$-plasmon frequency is from $5.1$ $eV$ to $12.0$ $eV$ and the $\sigma+\pi$ plasmon frequency centered at $14.5$ $eV$ for monolayer graphene. The threshold $\pi$-plasmon frequency increases with increasing stacking layer-number. The $\pi$-plasmon frequency of graphite begins at $\sim 7.1$ $eV$. Additionally, the monolayer graphere exhibits a linear dispersion of $\pi$-plasmons, as opposed to a parabolic dispersion for multilayer graphene. However, there was no detailed experimental investigation on $\pi$-plasmons invloving the stacking order and anisotropy of graphites. In fact, electronic excitations, except for the magnitude transferred momentum, also strongly depend on its direction, and the stacking order could further affect the anisotropy.

In this paper, we employ the tight-binding method to calculate the $\pi$
electronic band structures and use the self-consistent-field approach to
evaluate the dielectric function for AA-stacked graphite. The electronic
structure shows anisotropic characteristics in the hexagonal plane of the
first Brillouin zone. The anisotropy of $\pi$-band is reflected in special
features of the dielectric function and thereby the  loss function.
The interlayer interaction further enhances the anisotropy leading to the
remarkably distinct dispersion relations of $\omega_{p}$ versus $\textbf{q}$
between 2D graphene and 3D AA-stacked graphite.

The rest of this paper is organized as follows. A brief description of
the band structure is given in $\S$2. The results for our dielectric function,
plasmon dispersion relation and loss function are presented in $\S$3.
$\S$4 is devoted to  concluding remarks.

\section{Tight-binding Method and Electronic Structure}

For  AA-stacked graphite, the geometric structure is formed from periodically stacked monolayer graphene along the $z$-direction. Carbon atoms in each layer have the same projections on the hexagonal $x$-$y$ plane. The C-C bond length is  $b=1.42$ $\AA$, and the interlayer distance is $c=3.35$ $\AA$. Each hexagonal unit cell includes two atoms, and the atomic interactions $\gamma_{i}$  were obtained from the study of Charlier.$^{11,12)}$ The tight-binding model with $2p_z$ orbitals is employed to calculate the electronic structure of AA-stacked graphite. The band structure has been discussed in detail in  previous work.$^{23,45)}$ Before a further discussion on the properties of the electronic excitations, we briefly review  the important features of the band structure  of monolayer graphene and AA-stacked graphite which differ from those of AB-stacked and ABC-stacked graphite. This is done for completeness and for introducing our notation. The first BZ, shown in the inset of Fig. 1, includes the symmetry points $\Gamma\lbrack 0,0,0\rbrack$, $M\lbrack2\pi/(3b),0,0\rbrack$, $K\lbrack 2\pi/(3b),2\pi/(3\sqrt{3}b),0\rbrack$, $A\lbrack 0,0,\pi/c\rbrack$, $L\lbrack2\pi/(3b),0,\pi/c\rbrack$, and  $H\lbrack 2\pi/(3b),2\pi/(3\sqrt{3}b),\pi/c\rbrack$. The suffix indicates the
value of  $k_z$ (in units of   $\pi/(c)$ here and henceforth. The band
structure  along different symmetry directions is shown in Fig. 1.
.
\begin{figure}
\begin{center}
\includegraphics[width=15cm]{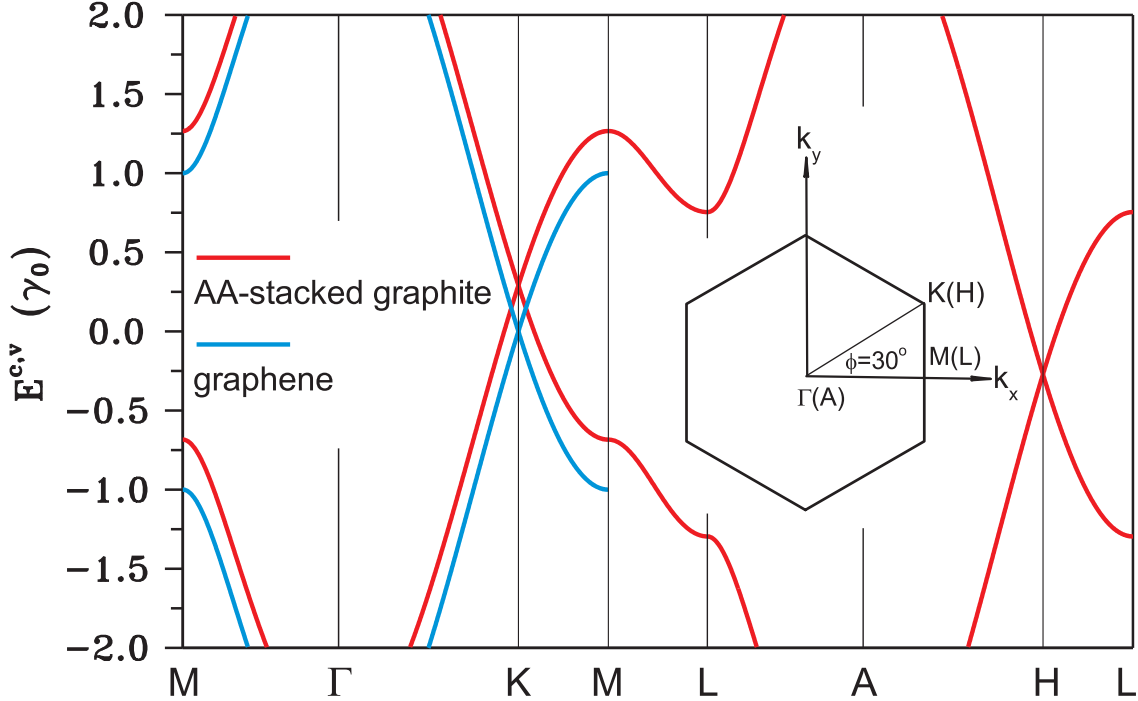}
\end{center}
\caption{(Color online) $\pi$-electronic structures of AA-stacked graphite and monolayer graphene are shown along symmetric directions. The first BZ in the inset.}
\label{f1}
\end{figure}

The $\pi$ energy bands of monolayer graphene are shown along $M\to \Gamma\to  K\to M$ in the first BZ in Fig. 1. The conduction and valence bands ($E^{c}$ and $E^{v}$)  are symmetric about the Fermi energy ($E_{F}=0$). Near the $K$ point, two linear energy bands intersect at the Fermi level and are isotropic  in $(k_x,k_y)$. At zero temperature, only interband transitions are allowed but  do not exhibit  low-frequency collective excitations. As the temperature is increased, the new allowed intraband transitions may induce low-frequency plasmons. The central-energy region, along $M\to \Gamma$ ($K\to M$) direction, has a local minimum (maximum) at the $M$ point with energy $E^{c,v}=\pm\gamma_{0} (\pm2.569$ eV). Energy dispersions around the $M$ point make it a saddle point, which induces the $\pi$-plasmon about $\omega_{p}\stackrel{>}{\sim} 2 \gamma_{0}$. As for AA-stacked graphite, the energy dispersions along $M\to \Gamma\to  K\to M$ ($k_{z}=0$) and $L\to A\to  H\to L$ ($k_{z}=\pi/c$) are similar to those of monolayer graphene. The Fermi energy is moved to 0.0208 eV, inducing the intraband excitations near the $K$ ($H$) points. Due to interlayer interactions, the band structure is shifted upward [downward] for $k_z=0$ [$k_z=\pi/c$] with respect to that of monolayer graphene, and the energy separation ($E^{s}=E^{c}-E^{v}$) is reduced [enlarged]. Additionally,  $E^{s}$ at the saddle points is increased as the states change from $M\to \Gamma$ and $L\to A$ ($\phi=0^o$), while they decrease as the states change from $M\to K$ and $L\to H$ ($\phi=90^o$ or $30^o$). The plasmons around $K$ point was studied in Ref. 22. Here, we focus on the $\pi$-plasmons. Those are excited  around the saddle points $M\to L$ in terms of the calculated band structure. We shall compare results for AA-stacked graphite and graphene. In the following section, we investigate anisotropy of $\pi-$plasmons and the effects due to interlayer interaction on them.

\section{Plasmon  Dispersion Relation}

\subsection{The  dielectric  function}

The $\pi$-electronic excitations, due to anisotropic $\pi$-bands,
are described by the transferred momentum
$\hbar\textbf{q} =(\hbar q_{\parallel} \cos \phi, \hbar q_{\parallel} \sin \phi, \hbar q_z)$ and the excitation energy $\hbar\omega$. Here, $\phi$ is
the angle between the in-plane transferred momentum $\textbf{q}_{||}$ and the
$\Gamma\to M$ direction. Since the first BZ is hexagonal,  the range
$0^{\circ}\le\phi\le 30^{\circ}$ is sufficient to characterize the direction-dependent excitations. Namely, both  $\phi$ and $\phi +60^{\circ}$
cases have the same electronic excitations spectra.

At arbitrary temperature $T$, the dielectric function calculated for bulk graphite in the RPA is$^{46)}$

\begin{eqnarray}
{\epsilon(\textbf{q},\phi,\omega)}=\epsilon_{0}-v_{q} \sum_{{h^\prime},h}{\int_{1stBZ}} 2{d^3{\bf k}\over\,(2\pi)^3} {|\langle{{\bf k}+{\bf q}; h^{\prime}}|e^{i{\bf q\cdot r}} |{\bf {k}}; h\rangle|^2} \frac{f(E_{h^{\prime}}({\bf {k+q}}))-f(E_{h}({\bf k}))} {E_{h^\prime}({\bf {k+q}})-E_h({\bf k})-(\omega+i\delta)} \  ,
\end{eqnarray}
where $\textbf{k}$  and $\textbf{q}$ are 3D wave vectors, $v_{q}=4\pi e^{2}/q^{2}$ is the bare Coulomb interaction, and $\epsilon_{0}=2.4$ is the background dielectric constant for graphite. $E_{h^\prime}({\bf {k+q}})$ and $E_{h}({\bf k})$ are the state energies of final and initial states and $h$ ($h^\prime$) labels conduction  or valence bands. $\delta= 0.1$ $\gamma_{0}$ is the energy width due to various deexcitation mechanisms and $f(E_h({\bf k}))$ is the Fermi-Dirac  distribution function. In demonstrating the anisotropy of the dispersion relation of $\pi$-plasmons, we first  consider inelastic scattering just involving $\textbf{q}$ along the hexagonal plane in the BZ, i.e., $k_{z}$ conserved and $q_{z}=0$. Then, we will turn to the case of finite $q_z$.

\par
The electronic excitations due to the critical points in the energy-wave-vector space would cause singular behavior in $\epsilon$. They include the $K$ point and the saddle point (near $M$). The imaginary (real) part $\epsilon_{2}$ ($\epsilon_{1}$), evaluated at the $K$ ($M$) point, exhibits a square-root divergence (logarithmic divergence). The $M$ point is related to the $\pi$-electronic excitations, while the $K$ point is mainly associated with the low-frequency ones. When $\delta$ approaches zero, the dip-like structure in $\epsilon_{1}$ becomes a discontinuity which accompanies zeros of $\epsilon_{1}$. In the long wavelength limit ($q\to 0$), the dip-like structure in $\epsilon_{1}$, due to vertical transition, occurs around $\omega\sim 2$ $\gamma_{0}$ (not shown here). With increasing $\delta$ and $q$, such a structure would become broadened and occurs at higher frequency. The vanishing $\epsilon_{1}$, if at which $\epsilon_{2}$ is small, is associated the $\pi$-plasmon. That is, when Landau damping is weak, EELS will exhibit a prominent plasmon peak.

The calculated $q$-dependent dielectric function of AA-stacked graphite is mainly considered at $\phi=0^{\circ}$ ($|k_{x},k_{y},k_{z}\rangle\to |k_{x}+q_{\parallel},k_{y},k_{z}\rangle$) and $\phi=30^{\circ}$ ($|k_{x},k_{y},k_{z}\rangle\to |k_{x}+q_{\parallel}\cos(30^{\circ}), k_{y}+q_{\parallel}\sin(30^{\circ}),k_{z}\rangle$).
At $\phi=0^{\circ}$ and ${q}_{||}=0.2$ (in units of  $\AA^{-1}$ here and hereafter), the real part $\epsilon_{1}$ (the red solid curve) and the imaginary part $\epsilon_{2}$ (the red dashed curve) are shown in Fig. 2(a). For $\epsilon_{2}$, single-particle excitations (SPE) exhibit two peaks at $\omega_{1}=0.436$ $\gamma_{0}$ and $\omega_{2}=2.201$ $\gamma_{0}$, respectively. The first peak ($\omega_{1}$) is due to intraband and interband transitions near the $K$ and $H$ points, while the second peak ($\omega_{2}$) only comes from interband transitions closely related to the $M$ and $L$ points.
While the frequency is in the vicinity of $\omega_{1}$, $\epsilon_{1}$ is changed from positive value into negative value that exhibits a dip-like structure. As $\omega$ is increased, the value of $\epsilon_{1}$ gradually becomes positive and has one zero at which $\epsilon_{2}$ vanishes. The zero value of  $\epsilon_{1}$ and the small $\epsilon_{2}$  lead to collective plasmon excitations, as shown below. As ${q}_{||}$ is increased, SPE of $\epsilon_{2}$ is mainly induced by interband excitations extended from the neighborhood of $K$ ($H$) point to $M$ ($L$) point in the BZ. Figures 2(b) and 2(c) show that $\epsilon_{2}$, except for the peak structure, also possesses a shoulder structure due to broadening. Additionally, the frequencies corresponding to the peak positions and the vanishing of $\epsilon_{2}$ increase as ${q}_{||}$ is increased. Simultaneously, the dip-like structure of $\epsilon_{1}$ is less significant and occurs at higher frequency as ${q}_{||}$ is increased. This result implies that the frequency and intensity of $\pi$-plasmons will have a  strong dependence on the magnitude of ${q}_{||}$.

\begin{figure}
\begin{center}
\includegraphics[width=15cm]{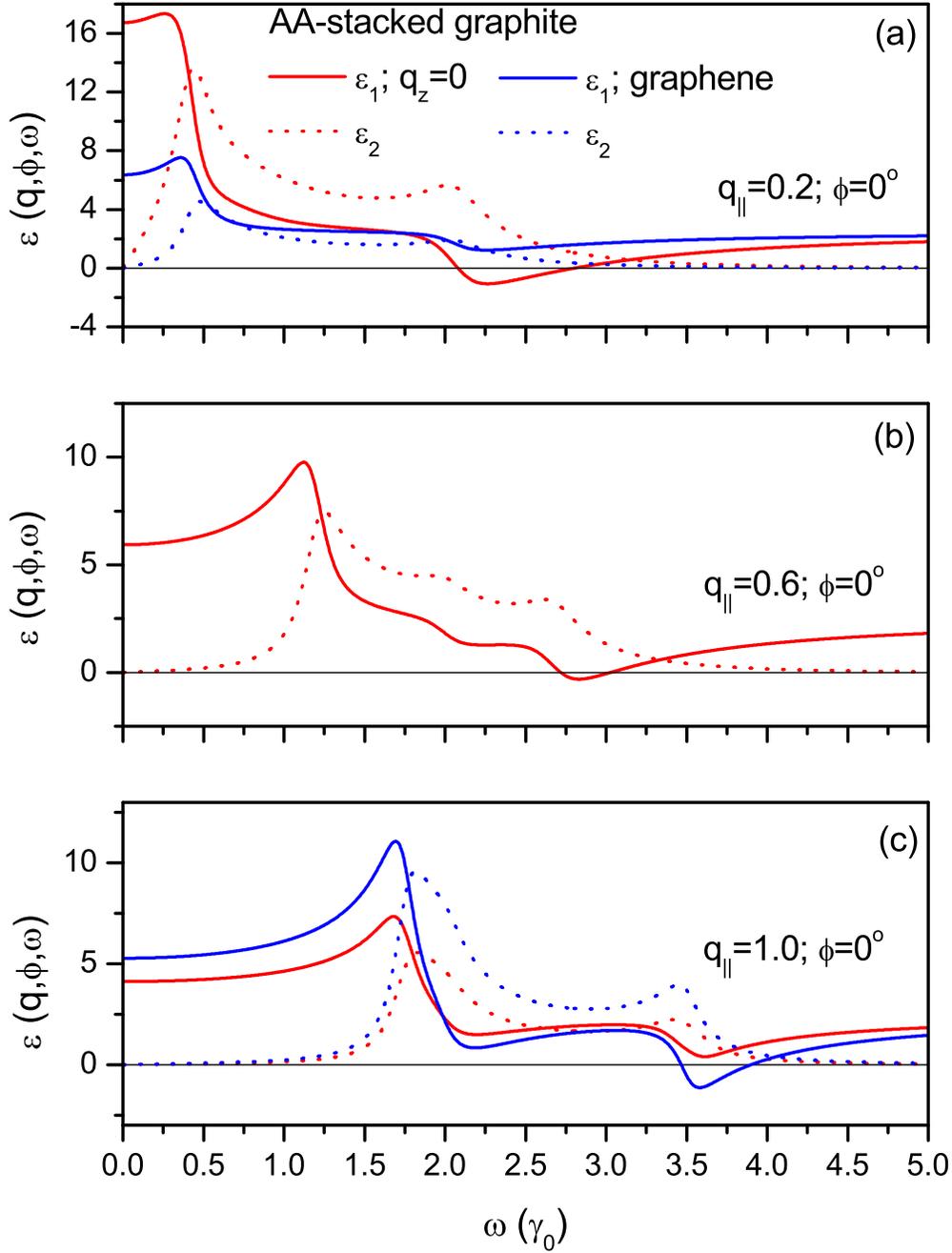}
\end{center}
\caption{ (Color online) The real ($\epsilon_1$) and imaginary ($\epsilon_2$) parts of the dielectric function  of AA-stacked graphite for $q_z=0,\ \phi=0^{\circ}$ are plotted as functions of frequency for fixed in-plane transferred momentum (a) $q_\parallel=0.2$, (b) $q_\parallel=0.6$, and $q_\parallel=1.0$. For comparison, the dielectric function of graphene is also plotted. At $q_\parallel=0.6$, both of them have the same $\epsilon$.}
\end{figure}

\begin{figure}
\begin{center}
\includegraphics[width=15cm]{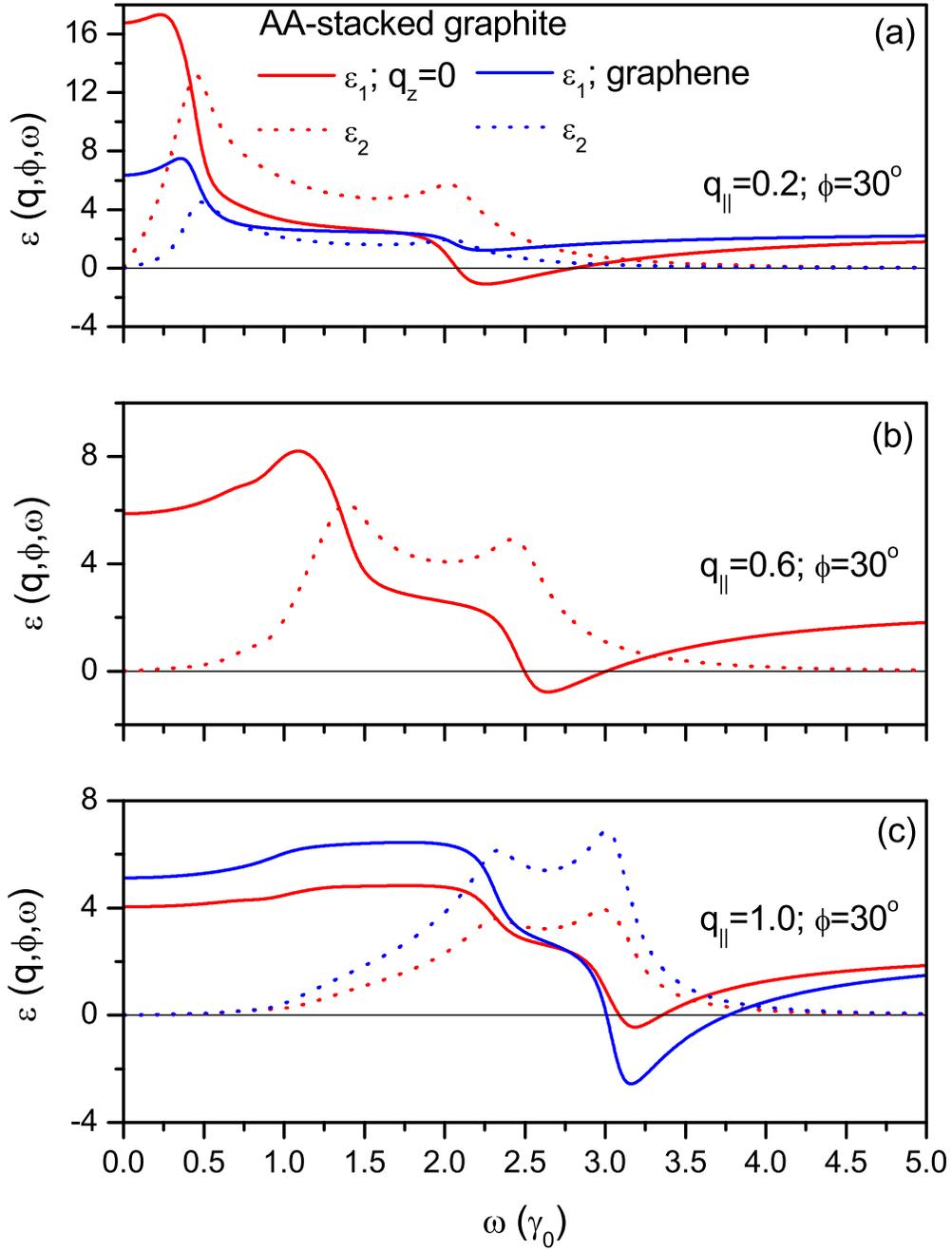}
\end{center}
\caption{ (Color online) The same plot as Fig. 2, except that $\phi=30^{\circ}$.}
\end{figure}

Apart from its dependence on the  magnitude of the transferred momentum, the dielectric function also depends on the  angle $\phi$. Figure 3 shows the ${q}_{||}$-dependent dielectric functions at $\phi=30^{\circ}$. It is found that $\epsilon$ has the same special structures for both $\phi=0^{\circ}$ and $\phi=30^{\circ}$ at small ${q}_{||}$, e.g., ${q}_{||}=0.2$. However, at larger ${q}_{||}$, peak structures ($\phi=0^{\circ}$) in $\epsilon$ are changed into nearly shoulder structures ($\phi=30^{\circ}$). The SPE energies at $\phi=0^{\circ}$ are larger than those at $\phi=30^{\circ}$, since the former (the later) are associated with $E^{s}$ along $M\to \Gamma$ and $L\to A$ ($M\to K$ and $L\to H$). Furthermore, $\epsilon_{1}$ shows more obvious dip-like structure and its zeros occur at smaller frequency. These results suggest that the spectra of plasmons have strong direction-dependence for larger transfer momentum. Furthermore, our results fully reflect the anisotropy of band structure for wave vectors far away from the saddle ($M$) and the corner ($K$) points.

In order to investigate  the effects due to interlayer interaction on $\epsilon$, Figs. 2 and 3 (blue curves)  also present the ${q}_{||}$-dependent dielectric function for monolayer graphene when  $\phi=0^{\circ}$ and $\phi=30^{\circ}$. Our results show that for small ${q}_{||}$, $\epsilon$ has a weak $\omega$-dependence on ${q}_{||}$. However, as ${q}_{||}$ is increased, some significant structures in $\epsilon$ are generated, leading to the same $\epsilon$ at $q=0.6$ for both graphene and AA-stacked graphite. Furthermore,  at larger ${q}_{||}$, e.g., ${q}_{||}=1.0$, $\epsilon$ for graphene has more  structure  than that of AA-stacked graphite. Form these results, it could be predicted that the frequency and the intensity of electronic excitations due to interlayer interactions will present very different ${q}_{||}$-dependence between graphene and AA-stacked graphite.

\subsection{The transferred-momentum-dependent loss spectra}

The loss function, defined as Im$[-1/\epsilon(\textbf{q},\omega)]$, is an indicator of the intensity of electronic excitations and is closely related to experimental
probes such as light and inelastic electron scattering spectroscopies. The function Im$[-1/\epsilon]$ is plotted in Fig. 4(a) for $\phi=0^{\circ}$,
$q_z=0$  and $\delta=0.1$ $\gamma_{0}$ for several values of  $q_\parallel$. The  peak above $2.8$ $\gamma_{0}$ corresponds to the interband $\pi$-plasmon. The shoulder structure occurring at lower frequency originates from the $\pi\to\pi^{\ast}$ single-particle excitations. The plasmon frequency increases with the transferred momentum, while the intensity of the plasmon peak is decreased as $q_\parallel$ is increased.   $\epsilon_{1}$ approaches zero more gradually and the singular structures in $\epsilon$ become less pronounced (shown in Fig. 2(a)) is the main reason, since the derivative of $\epsilon_{1}$ versus $\omega$ is inversely proportional to the strength of collective electronic excitations. The loss spectra as well as dielectric functions are  very sensitive to changes in the direction of the transferred momentum. Plasmons at $\phi=30^{\circ}$ are distinctively different from those at $\phi=0^{\circ}$, the plasmon frequency increases more slowly with $q_\parallel$ and its intensity seems to be ${q}_{||}$-independent, as shown in Fig. 4(b). Compared with the 3D AA-stacked graphite, 2D graphene exhibits very different ${q}_{||}$-dependent loss functions, as shown in Fig. 4(c) for $\phi=0^{\circ}$. With increasing $q_\parallel$, both the plasmon frequency and its intensity for graphene increase more rapidly than those of AA-stacked graphite. In short, the high anisotropy of the electronic structure in the hexagonal plane of the BZ and the interlayer interactions are directly reflected in the dielectric functions and the frequency and  intensity of the plasmon spectra.

\begin{figure}
\begin{center}
\includegraphics[width=15cm]{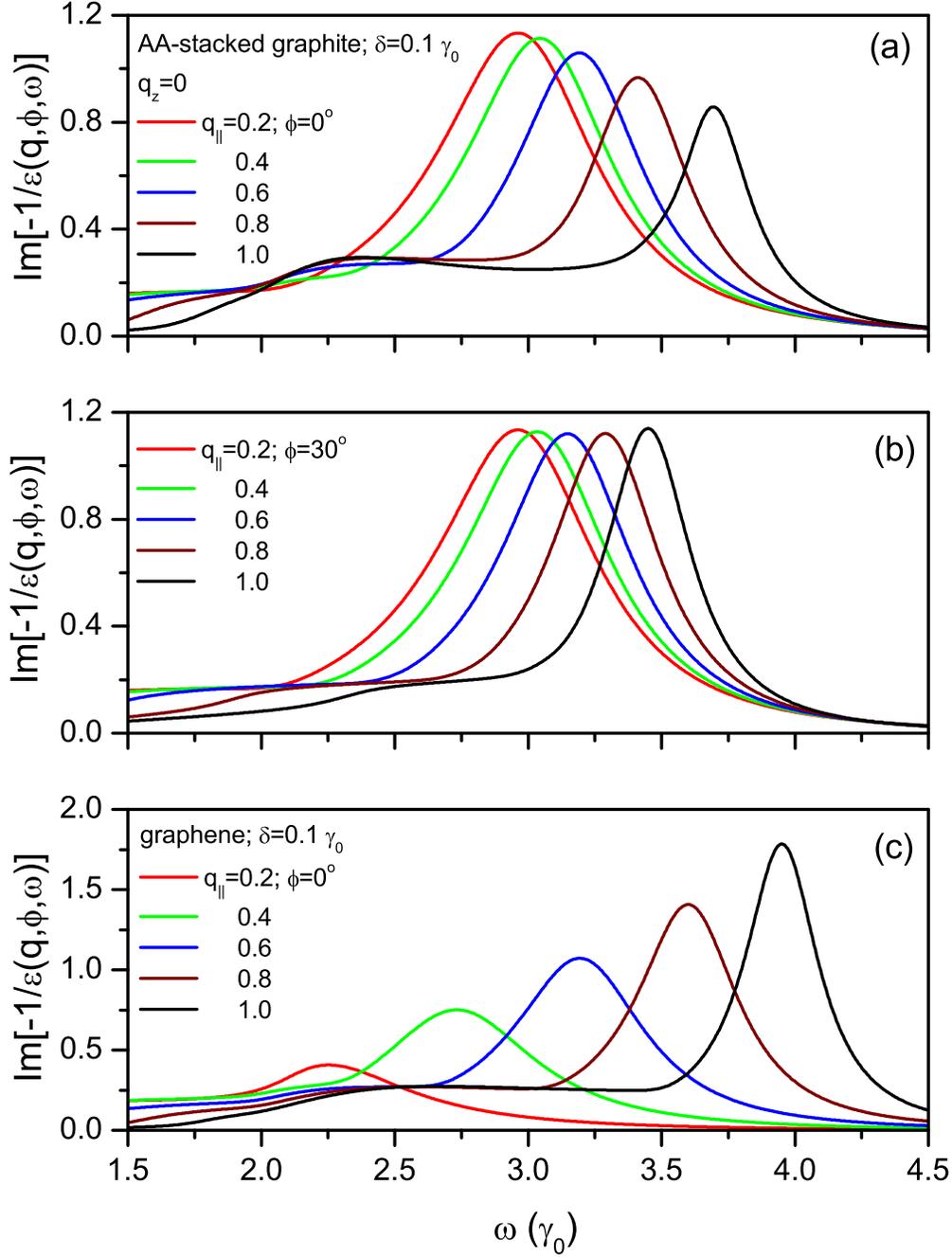}
\end{center}
\caption{(Color online) The energy loss spectra of AA-stacked graphite for several chosen $q_\parallel$ at (a) $\phi=0^{\circ}$ and (b) $\phi=30^{\circ}$. The loss spectra of graphene for different $q_\parallel$ at $\phi=0^{\circ}$ are plotted in panel  (c).}
\end{figure}

\begin{figure}
\begin{center}
\includegraphics[width=15cm]{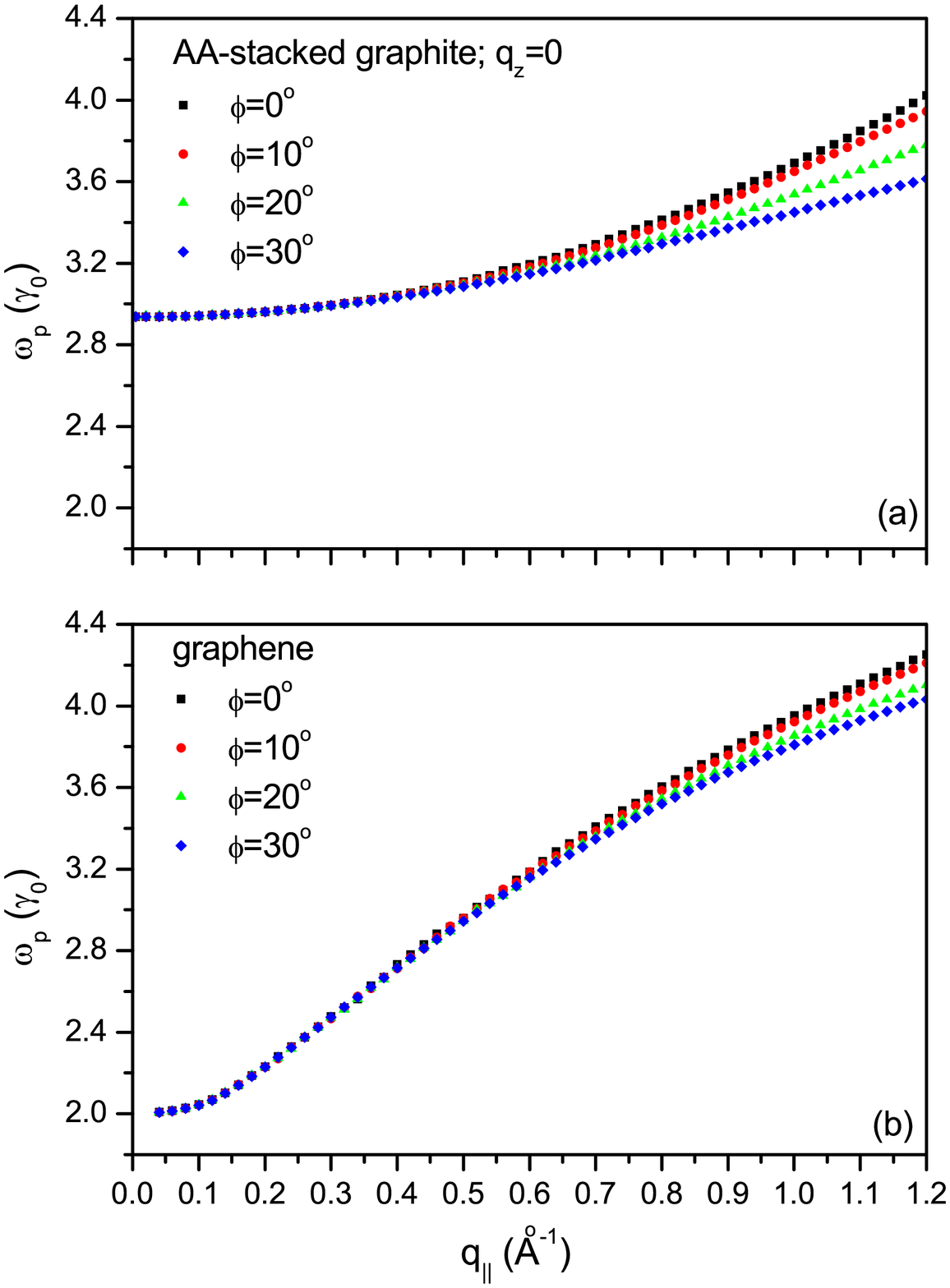}
\end{center}
\caption{(Color online) The $q_\parallel$-dependent $\pi$-plasmon frequency is presented  for several chosen angles $\phi$ for (a) AA-stacked graphite and (b) monolayer graphene.}
\end{figure}

The $q_\parallel$ and $q_z$-dependent behavior of the $\pi$-plasmon is important in understanding the $\pi$-band features in graphite. The strong dispersion relation of the plasmon frequency with $q$ means that the plasma oscillation behaves as a propagating wave with wavelength $2\pi/q$ and group velocity $\nabla_q\omega_{p}(q)$. The dispersion relation of the $\pi$-plasmon frequency with $q_\parallel$ is shown in Fig. 5(a) for $q_z=0$ and different angles $\phi$. The plasma frequency $\omega_{p}$ ($\sim 2.9$ $\gamma_{0}=7.4$ eV) is finite when $q_\parallel\to 0$ and is within the region of  optical scattering
spectroscopies. Therefore, $\pi$-plasmon is an  optical plasmon. Our results show that the plasmon frequency exhibits a strong ${q}_{||}$ dependence, and that the ${q}_{||}$-dependence of $\omega_{p}$ is gradually reduced with increasing $\phi$. This result directly reflects the $\pi$-band characteristics. We further fit the dispersion relations with the quadratic equation $\omega_{p}=B_{\circ}+ B_{1}q+B_{2}q^{2}$. $B_{\circ}\sim 2.92$ corresponds to the threshold $\pi$-plasmon frequency for ${q}_{||}\to 0$. By varying $\phi$ form $\phi=0^{\circ}$ to $\phi=30^{\circ}$, $B_{2}$ decreases form $B_{2}=0.793$ to $B_{2}=0.348$, while $B_{1}$ increases form $B_{1}=-0.043$ to $B_{1}=0.181$. The fitting result reveals that the dispersion relation at small $\phi$ presents a quadratic feature. However, the quadratic form is gradually reduced into the nearly linear form with increasing $\phi$. The anisotropic band structures in Fig. 1 could explain the above result, since $\phi=0^{\circ}$ and $\phi=30^{\circ}$ respectively represent ${q}_{||}$ along $\Gamma\to M$ and $\Gamma\to K$

 Figure 5(b) reveals that the $\pi$-plasmons of graphene are also optical
 modes.$^{47)}$ However, their $\omega_{p}$-${q}_{||}$ dispersion relation differs from
 that for the $\pi$-plasmons of AA-stacked graphite. For example, for
 small  momentum transfer, the $\pi$-plasmon frequency approaches
 $\omega_{p}\sim 2.0$ $\gamma_{0}$ (=5.1 eV). The threshold frequency
is close to the  maximum single-particle excitation frequency from the $M$ point
and much smaller than that for AA-stacked graphite. This  means that the
depolarization shift in graphite is larger than graphene because the screening
of the Coulomb interaction is larger in the 3D system compared to that in 2D.
Our calculations show that for $0\le {q}_{||}\le 1.2$, the width of the plasmon
 excitation region for $\phi=0^{\circ}$ is  $\Delta\omega_{p}\approx 1.09$ $\gamma_{0}$
 for AA-stacked graphite, but it is
$\Delta\omega_{p}\approx 2.25$ $\gamma_{0}$ for graphene.
That is, the plasmon dispersion relation for graphene displays stronger
$q$-dependence than that of AA-stacked graphite. However, it has also
been found that as $q$ is increased, $\omega_{p}$ significantly shows
a stronger $\phi$ dependence for AA-stacked graphite than for monolayer graphene.
Specifically, at ${q}_{||}=1.2$ the difference in plasmon frequency between
$\phi=0^{\circ}$ and $\phi=30^{\circ}$ is $\Delta\omega_{p}\approx 0.409$
$\gamma_{0}$ for AA-stacked graphite, but it is $\Delta\omega_{p}\approx 0.223$
$\gamma_{0}$ for graphene. Additionally, our curve fitting results for
various values of $\phi$  show strong linearity for the  dispersion relations  of graphene.
We deduce from these results  that interlayer interactions significantly
enhance the anisotropy of in-plane $\pi$-plasmons, which is unusual for
optical excitations.

\begin{figure}
\begin{center}
\includegraphics[width=15cm]{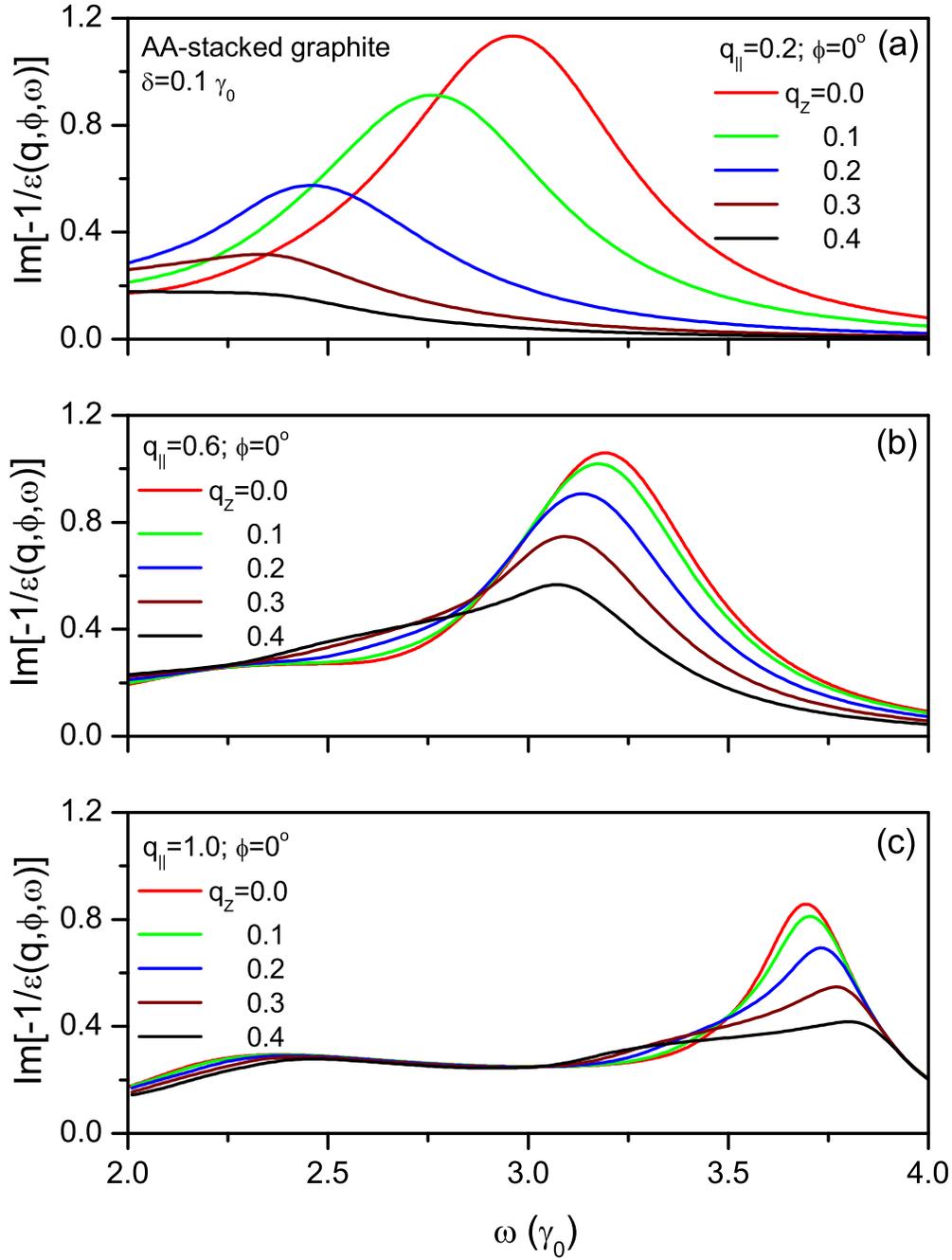}
\end{center}
\caption{(Color online) The energy loss spectra of AA-stacked graphite for $\phi=0^{\circ}$ and chosen in-plane transferred momentum  (a) $q_\parallel=0.2$,  (b) $q_\parallel=0.6$ and (c) $q_\parallel=1.0$.  As $q_z$ is increased, the sharpness of the resonance peak is diminished until there is no plasmon resonance, indicating a cut-off, for fixed $q_\parallel$, as shown in Fig. 7.}
\end{figure}

\begin{figure}
\begin{center}
\includegraphics[width=15cm]{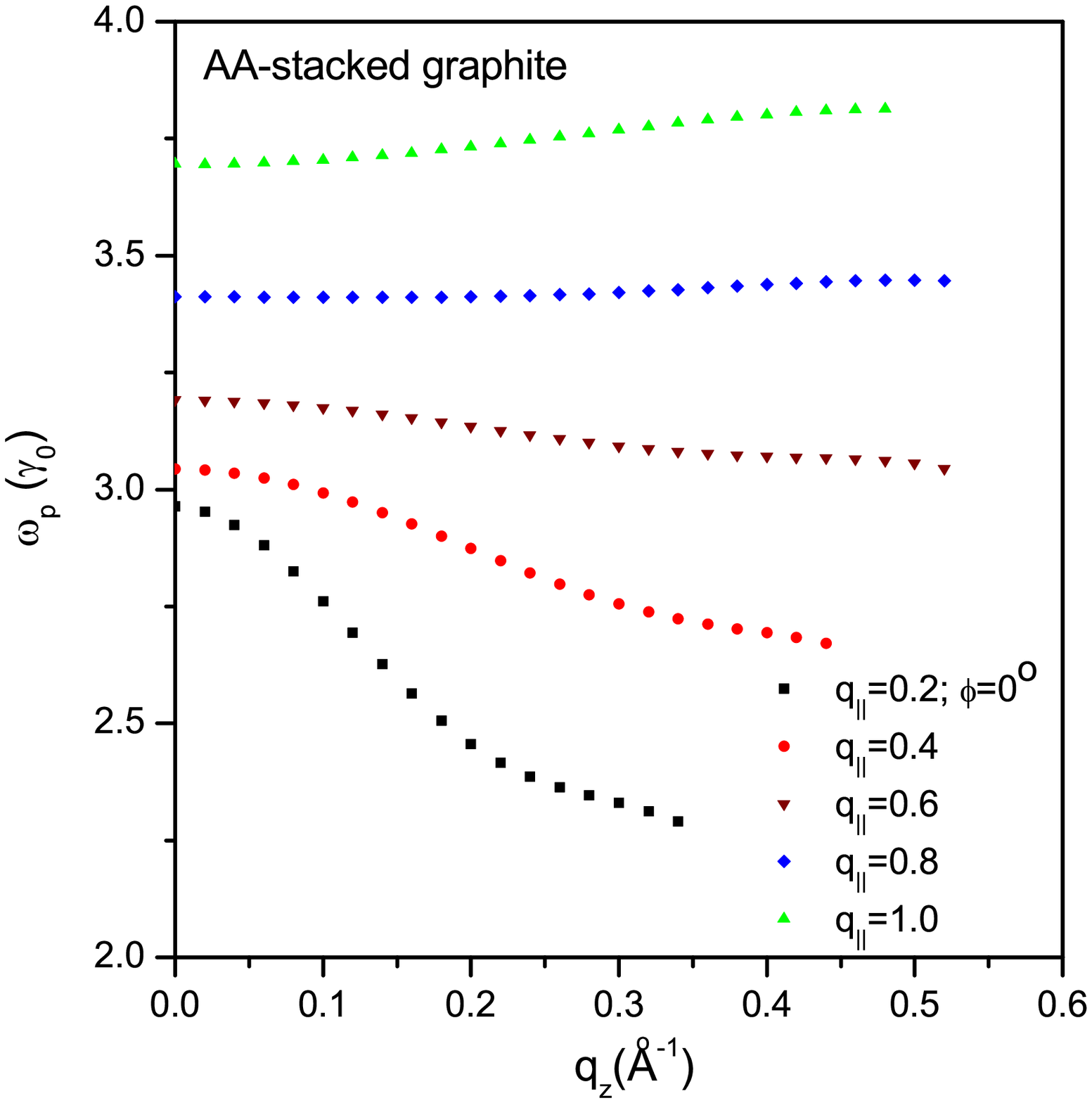}
\end{center}
\caption{(Color online) The dispersion relation for $\pi$-plasmons of AA-stacked graphite as a function of $q_z$ for  $\phi=0^\circ$ and fixed ${q}_\parallel$. Each plasmon branch has an upper cut-off $q_z$, above which there is no $\pi$  plasmon. The plasmons cease to exist beyond a cut-off $q_z$ which depends on the chosen $\textbf{q}_\parallel$.}
\end{figure}

In Fig. 6,  the loss function   for $\pi$ plasmons in
 AA-stacked graphite  is plotted versus frequency  for $\phi=0^{\circ}$
and chosen in-plane transferred momentum  (a) $q_\parallel=0.2$,  (b) $q_\parallel=0.6$ and
(c) $q_\parallel=1.0$ and $\delta=0.1$ $\gamma_{0}$ for several values of  $q_z$.
As $q_z$ is increased, the sharpness of the resonance peak is
reduced until there is no plasmon observable plasmon resonance.
All the  peaks appear  above $2.0$ $\gamma_{0}$ which corresponds to the
single-particle interband transition energy near  the shoulder  for the
$\pi\to\pi^{\ast}$ transitions. The plasmon frequency increases with the
transferred momentum $q_z$ for fixed large $q_\parallel$, but decreases with increasing $q_z$ for small $q_\parallel$. Additionally, the intensity
of the plasmon peak is decreased as $q_z$ is increased.
Clearly, the loss spectra  are  very sensitive to changes
in the magnitude of the perpendicular transferred momentum.
 Plasmons at large $q_\parallel$
are crucially different from those at small $q_\parallel$,
since their group velocities would have different signs. These results
show that compared with the $q_\parallel$ dependence, the $q_z$ dependence
of the loss function is very much different for AA-stacked graphite.
Furthermore, while there is a lower bound for the in-plane transferred momentum,
there is an upper bound for the perpendicular component of the
transferred momentum for $\pi$ plasmons, further illustrating their anisotropy.

In Fig. 7, the $\pi$-plasmon dispersion relation of  AA-stacked graphite
is presented as a function of $q_z$.   For sufficiently large $q_z$,
e.g., $q_z\stackrel{>}{\sim} 0.53$, this $\pi$-band induced plasmon barely exists
when the in-plane transferred momentum $q_\parallel=1.0$. In this case, the plasmons
are Landau damped by the single-particle excitations. For small $q_z$, the
group velocity for any of the $\pi$-plasmon branches depicted in Fig. 7 is very small. However, its magnitude generally increases, but can be positive
or negative depending on the value of $q_\parallel$.
The anisotropy of the band structure in the hexagonal plane of the
BZ and the interlayer interactions are  again reflected in the dielectric functions
and the frequency and  intensity of the plasmon spectra.

Both the stacking sequence and the dimensionality play an important role in the electronic excitations. The high-energy plasmons are reported for the 2D single-layer graphene and multilayer graphene with the AB and ABC stacking in a recent study by Yuan et al. by means of similar calculated theories.$^{48)}$ The unique features in the excitation spectra due to the stacking sequence are mainly from the excitations of the states near the M points, but they exclude the results from the Dirac Cone Approximation. Furthermore, the more graphene layers there are the more complicated electronic structures become, and it is the same for plasmons. Comparison with 3D graphite, the $q_z$ dependence of plasmons is absent for 2D systems. However, the dependence on the number of layer exhibits the interesting and rich results. They are expected to exist in the AA-stacked multilayer graphene. In a nutshell, the dependences of plasmons on the number of layer, $q_z$, the stacking sequence, and the dimensionality provide clear forms of the electronic excitations needed for the experiments and are also conducive to further understanding graphene systems.

\section{Concluding Remarks}

In summary, in this paper, we used the tight-binding method to calculate the $\pi$
electron band structure of AA-stacked graphite. The band structure is
highly anisotropic in the hexagonal plane of the BZ. But, it has a weak dependence
on $k_{z}$ and has a narrower band width along the $k_{z}$ direction. The
dielectric function is evaluated in the self-consistent-field approach.
The $\pi$-plasmons are optical plasmons and show isotropic behavior  for small
transferred momentum. As the transferred momentum is increased, the anisotropy
of the $\pi$-band is transformed  into features appearing in  $\epsilon$,
 including a dip-like structure in $\epsilon_{1}$ and a peak structure in
 $\epsilon_{2}$. The special structures are significantly altered by changing
  the magnitude and direction of the transferred momentum.
Therefore, the plasmon frequency and the peak position of the loss functions
also show strong $q_\parallel$- and $\phi$-dependence. There exist many significant
 differences for the $\pi$-plasmon dispersion relations between 2D graphene
 and 3D AA-stacked graphite.
For the former, the plasmon
 frequency has a stronger $q_\parallel$-dependence, resulting in a wider
$\pi$-plasmon band. However, $\phi$-dependence of $\pi$-plasmons, for
graphite, becomes stronger with increasing $q_\parallel$. These results fully reveal the
role of interlayer interaction to enhance the anisotropy of
$\pi$-plasmons.  We have also presented the plasmon dispersion relation as a function
of $q_z$ for fixed  $q_\parallel $. We  show that there is an upper bound on
$q_z$ for plasmons to exist in graphite. Furthermore, the group velocity for
plasmon propagation along $q_z$ may be positive or negative depending
on the value of $q_\parallel$, as we demonstrate in Fig. 7.
Our derived results could be verified by measurements
of electron-energy-loss spectra or optical spectra.








\end{document}